\documentclass[12pt,preprint]{aastex}

\def\deg{$^{\rm o}$}
\def\ltsim{\raisebox{-.5ex}{$\;\stackrel{<}{\sim}\;$}}
\def\gtsim{\raisebox{-.5ex}{$\;\stackrel{>}{\sim}\;$}}

\shortauthors{Crenshaw \& Kraemer}
\shorttitle{Absorption in Reddened Seyfert 1 Galaxies}

\begin{document}

\title{On the Nature of Intrinsic Absorption in Reddened Seyfert 1 Galaxies}

\author{D.M. Crenshaw}
\affil{Department of Physics and Astronomy, Georgia State University,
Atlanta, GA 30303; crenshaw@chara.gsu.edu}

\author{S. B. Kraemer}
\affil{Catholic University of America and Laboratory for Astronomy and
Solar Physics, NASA's Goddard Space Flight Center, Code 681,
Greenbelt, MD  20771; stiskraemer@yancey.gsfc.nasa.gov}

\begin{abstract}

We discuss the origin of the ``dusty lukewarm absorber'', which we previously 
identified in the reddened Seyfert 1 galaxies NGC~3227 and Akn~564. This 
absorber is characterized by saturated UV absorption lines (C~IV, N~V) near 
the systemic velocity of the host galaxy, and is likely responsible for 
reddening both the continuum and the emission lines (including those from the 
narrow-line region) from these Seyferts. From a large sample of Seyfert 1 
galaxies, we find that continuum reddening (as measured by UV color) tends to 
increase with inclination of the host galaxy. Furthermore, reddened, inclined 
Seyfert galaxies observed at moderate to high spectral resolution all show 
evidence for dusty lukewarm absorbers. We suggest that these absorbers lie in 
the plane of the host galaxy at distances $\gtsim$~100 pc from the nucleus, 
and are physically distinct from the majority of intrinsic absorbers that are 
outflowing from the nucleus. 

\end{abstract}

~~~~~

\section{Introduction}

A large fraction of Seyfert 1 galaxies show significant reddening of their 
continua and emission lines due to an intrinsic column of dust (Ward et al. 
1987). X-ray observations of a number of these galaxies show that the column 
density of neutral hydrogen is much smaller than that derived from the 
reddening 
(Komossa et al. 2000, and references therein). Brandt, Fabian, \& Pounds 
(1996) 
suggested that the dust exists within a warm absorber, characterized 
by O~VII and O~VIII bound-free edges, although they mentioned the possibility 
that the dust exists in lower ionization gas in which the hydrogen is still 
stripped. 
In Kraemer et al. (2000), we found that lower ionization gas in the form of a 
dusty ``lukewarm absorber'' provides a more natural explanation than the dusty 
warm absorber,
because it can be placed at a sufficient radial distance from the nucleus 
($\gtsim$ 100 pc) to redden most of the emission from the narrow-line region 
(NLR). We predicted that the lukewarm absorber could be identified by strong, 
and in some cases saturated, UV absorption lines (C~IV, N~V, Si~IV, and Mg~II)
in reddened Seyfert 1 galaxies.

We subsequently identifed lukewarm absorbers in {\it Hubble Space 
Telescope} ({\it HST}) spectra of the Seyfert 1 galaxy NGC~3227 (Crenshaw et 
al. 
2001a) and the narrow-line Seyfert 1 galaxy Akn 564 (Crenshaw et al. 2001b).
In both cases, photoionization models of the UV absorption lines confirmed 
that 
the column of gas (N$_H$ $=$ 1 -- 2 x 10$^{21}$ cm$^{-2}$) is sufficient to 
provide the observed reddening, assuming a dust-to-gas ratio that is within a 
factor of two of the Galactic value given by N(H) $=$ 5.2 x 10$^{21}$ E(B$-$V) 
cm$^{-2}$ (Shull \& van Steenberg 1985).
Interestingly, the galactic disks of both Seyferts are inclined ($i =$ 63\deg\ 
for NGC~3227, $i =$ 59\deg\ for Akn 564; de Zotti \& 
Gaskell 1985), so that our line of sight intercepts more gas and dust in their 
disks than for face-on galaxies. In addition, the intrinsic 
absorbers for both objects have 
velocities within 200 km s$^{-1}$ of the systemic velocities of their host 
galaxies, consistent with an origin in the galactic disks (Unger et al. 1987), 
whereas most UV absorbers show substantial outflow velocities with 
respect to their nuclei (up to $-$2100 km~s$^{-1}$, see Crenshaw et al. 1999). 
Finally, the depths of the UV lines confirm that the absorbers cover the NLRs 
in 
these objects, which indicates that they are at distances 
$\gtsim$ 100 pc from the Seyfert nuclei (Crenshaw et al. 2001a, 2001b). These 
findings led us to suggest that the dusty lukewarm absorbers lie in the 
galactic 
disks of these Seyferts, and may due to a highly ionized component in the 
extended narrow-line region (ENLR). In this letter, we pursue this suggestion 
by 
investigating the connection between reddening, UV absorption, and host galaxy 
inclination in a large sample of Seyfert galaxies.

\section{The Sample}

We have assembled data from the literature and the {\it IUE} and {\it HST} 
archives for 48 Seyfert 1 galaxies; these data are listed in Table 1. The list 
includes the 30 Seyfert 1 galaxies in the Ward et al. (1987) study of 
continuum 
reddening, which contains the complete sample of hard X-ray selected sources 
of 
Piccinotti et al. (1982) plus a few other Seyferts. Four objects from the Ward 
et al. sample were excluded from our study (NGC~2992, NGC 7172, and NGC 7582 
are 
generally considered to be Seyfert 2 galaxies, and 3C~273 is a quasar). We 
have 
also included Seyfert 1s in our sample that appear in recent studies of UV 
absorption (e.g., Ulrich 1988; Crenshaw et al. 1999) or X-ray absorption 
(e.g., 
Reynolds 1997; George et al. 1998).

Ward et al. (1987) found that at least 30\% of the Seyfert 1 galaxies in their 
sample showed signs of heavy reddening. Their broad-band 1 - 20 $\mu$m 
photometry combined with previous optical and IRAS fluxes indicated that the 
continuum energy distributions of these objects could be placed into one of 
three categories: A) bare, minimally reddened, B) reddened, or C) contaminated 
by the host galaxy. Seyferts with strong blue fluxes (i.e., at 3600 \AA) were 
placed in class A, those with steep visible spectra (decreasing to the blue) 
and 
flat far-infrared spectra were placed in class B, and those with steeply 
rising 
flux from 12 to 60 $\mu$m were placed in class C. We list these designations 
in 
Table 1.

We also list the axial ratios (b/a) of the host galaxies from, in order 
of preference, de Zotti \& Gaskell (1985), Kirhakos \& Steiner (1990), 
NASA/IPAC 
Extragalactic Database (NED), or direct measurement from the Palomar Sky 
Survey 
(POSS) images, as noted. We prefer the de Zotti \& Gaskell (1985) compilation, 
since it uses measurements from large-scale plates when available; the other 
measurements rely strictly on POSS images.
In Table 1, we note the presence or absence of UV absorption based on 
{\it HST} spectra obtained at moderate to high spectral resolution 
($\lambda$/$\Delta\lambda$ $\gtsim$ 1000). Most of these are from Crenshaw et 
al. (1999), although results on the following individual objects are also 
included: Ton~S180 (Turner et al. 2001), NGC 3227 (Crenshaw et al. 2001a), and 
NGC 4051 (Collinge et al. 2001). We also include a Seyfert 1 galaxy that shows 
significant evidence for intrinsic C~IV and Mg~II absorption in {\it IUE} 
spectra: MCG 8-11-11 (Ulrich 1988). We list the Galactic reddening E(B$-$V) 
values from NED that were derived from the maps of Schlegel et al. (1998).

In order to estimate the extinction, we determined a UV color from the 
continuum 
fluxes in {\it HST} or {\it IUE} spectra. For those sources with {\it HST} 
spectra, we used our measurements of continuum fluxes at 1350 \AA\ 
(far-UV) and 2200 \AA\ (near-UV). These wavelengths were chosen to avoid 
contamination by line 
emission; in particular,  the 2200 \AA\ region is just shortward of the 
``little 
blue bump'', which consists of Fe~II emission in the UV (Wills, Netzer, \& 
Wills 
1985). Unfortunately,
the sensitivity of {\it IUE} at 2200 \AA\ is quite poor; thus, for
objects with only archival {\it IUE} spectra, we
used the region around 2400~\AA\ for the near-UV point (we discuss the effects 
of this choice in \S 3). We obtained the {\it IUE} 
fluxes from the tabulated values in the AGN compilation of 
Courvoisier \& Paltani (1992); in cases of multiple observations, we used the 
first pair of well-exposed SWP and LWP images obtained on the same day. A few 
objects were observed after the publication 
of this catalog, and their fluxes were taken directly from spectra in the 
{\it IUE} archives. Note that we do not
regard this as a complete sample, since the requirement for UV spectra
biases the sample towards Seyferts with significant UV flux. However, the 
sample is large enough to show clear trends among the reddening properties, 
as we will demonstrate.

\section{Reddening, Inclination, and Absorption}

To investigate the trends in our sample, we use the inclination of the 
host galaxy:
\begin{equation}
i = cos^{-1}(b/a)
\end{equation}
and define a UV color (in magnitudes):
\begin{equation}
FUV-NUV = -2.5~log~[Flux(far\!-\!UV)/Flux(near\!-\!UV)].
\end{equation}

In Figure 1, we plot these two quantities for Seyferts designated type A or B 
by 
Ward et al. (we omit the point for 3C~120, since its type B classification is 
due to high Galactic reddening; see Table 1). The UV colors have been 
corrected 
for Galactic reddening using the curve of Savage \& Mathis (1979).

Figure 1 shows that the UV color FUV$-$NUV provides a measurement of 
reddening that is consistent with the Ward et al. results; those objects with 
FUV$-$NUV $<$ $-$0.2 are type A (unreddened) and those with FUV$-$NUV $>$ 
$-$0.2 
are type B (reddened). Thus, we can use the UV color as a consistent reddening 
indicator for the Seyferts in our entire sample. 
We note that this correlation implies that the reddening curves of Seyfert 
galaxies typically lack a significant 2200 \AA\ bump, since the bump yields
comparable extinctions at 1350 \AA\ and 2200 \AA\ in the standard Galactic 
curve (Savage \& Mathis 1979).
This conclusion is consistent with the lack of a 2200 \AA\ bump in the 
reddening 
curves of the Seyfert galaxies NGC 3227 (Crenshaw et al. 2001a), Akn 564 
(Crenshaw et al. 2001b), and quasars in general (Pitman, Clayton, \& Gordon 
2000).

Figure 1 also demonstrates that the inclined sources tend to be the most 
reddened, which indicates that the reddening is primarily associated with dust 
in the plane of the host galaxy. Similar correlations 
have been found in previous studies. Cheng, Danese, \& de Zotti (1983) found 
that the optical colors B$-$V and U$-$B of Seyfert 1 nuclei increase (i.e, 
become redder) with decreasing axial ratio (i.e., increasing inclination). De 
Zotti \& Gaskell (1985) found that the Balmer decrement H$\alpha$/H$\beta$ of 
the broad lines increases with decreasing axial ratio, and the narrow-line 
decrement shows a similar, though less significant, behavior. We point out the 
advantage of using the UV color for this type of correlation is that it is 
very 
sensitive to reddening, since reddening curves tend to be steep in the UV.

In Figure 2, we have replotted the UV color against inclination angle
for the entire sample, flagging those sources for which the presence or
absence of intrinsic UV absorption is known. The correlation between 
UV color and inclination is well established in this sample.
The dispersion in UV colors at low inclinations is due in large part to the 
intrinsic dispersion in spectral energy distributions, since the UV color of 
even an individual Seyfert can vary significantly. For example, based on the 
{\it IUE} fluxes of Fairall 9 in Courvoisier \& Paltani (1992), FUV $-$ NUV 
varied from $-$0.3 to $-$1.0 as it went through a factor of $\sim$30 variation 
in flux at 1350 \AA\ (with more negative colors corresponding to larger 
fluxes).
Other sources of dispersion at all inclinations could be: 1) different amounts 
of dust in the lines of sight through the host galaxies (due to intrinsic 
variations in dust content or clumpiness),  2) different reddening curves (see 
Crenshaw et al. 2001b),  or 3)  additional components of dusty gas that are 
not coplaner with the host galaxy (see \S 4). The use of the 2400 \AA\ region 
(instead of 2200 \AA) for the {\it IUE} spectra introduces only a small amount 
of scatter. From the STIS spectrum of NGC~4151 , which has a relatively large 
``little blue bump" (Crenshaw et al. 2001a), we find that the Fe~II 
emission in this region increases FUV$-$NUV by only $+$0.10, whereas the use 
of the continuum fit at 2400 \AA\  (i.e., if NGC~4151 had no emission bump) 
decreases FUV$-$NUV by only $-$0.09.

The Seyfert 1 galaxies in Figure 1 with intrinsic absorption are distributed 
thoughout the full range of inclination angle. This is not surprising, since 
many of these absorbers are known to be associated with the active nuclei (and 
not the host galaxies), due to their high outflow velocities, 
variability, or less than one covering factors (Crenshaw et al. 1999). 
However, it is interesting that absorption-free Seyfert 1 galaxies are 
completely lacking at high inclination angles. We suggest that the reason for 
this is that sources in this area of the plot tend to be reddened,  and a 
column of dusty gas which causes even 
marginally detectable reddening will produce detectable UV absorption lines, 
unless very highly ionized. For example, we have 
generated a set of models using the photoionization code CLOUDY90 (Ferland et 
al.1998) and the form of the ionizing continuum described in Kraemer et al. 
(2000). We varied the ionization parameter (number of ionizing photons per 
hydrogen atom at the illuminated face of the slab) for a fixed total hydrogen 
column of 5 x 10$^{20}$ cm$^{-2}$, which corresponds to a reddening of 
E(B$-$V) 
$=$ 0.10 (Shull \& van Steenberg 1985). We asssumed solar abundances 
(see Grevesse \& Anders 1989), and a 65\% depletion of the carbon onto grains. 
We find that for ionization parameters U $<$ 3.0, the column density of C~IV 
will be $>$ 10$^{13}$ cm$^{-2}$, which is the approximate detection limit for 
current high-resolution STIS spectra (Kraemer et al. 2001).

The labeled Seyfert 1 galaxies in Figure 2 are particularly interesting. All 
show evidence for reddening, and in addition, saturated C~IV and N~V 
absorption 
lines near the systemic velocities of their host galaxies. As we have shown, 
these are the signatures of dusty lukewarm absorbers in NGC~3227 and Akn~564, 
and we expect that future high-resolution spectra and modeling of the 
absorbers in WPVS 007 and MCG 8-11-11 will show that the columns of UV 
absorbing 
gas are sufficient to provide the observed reddenings.
Furthermore, we predict that the other heavily reddened, inclined Seyfert 
galaxies (NGC~931, IC~4329A, NGC~5506, and MCG -6-30-15 in particular) will 
show 
dusty lukewarm absorbers.

\section{Discussion}

We conclude that there are at least two types of UV absorbers in Seyfert 1 
galaxies: dusty lukewarm absorbers in the plane of the host galaxy and 
absorbers intrinsic to the nucleus. The first type is a natural consequence of 
viewing the nucleus through a host galaxy at a large inclination angle. It can 
be identified through saturated UV absorption lines (typically C~IV and N~V) 
near the systemic velocity in a Seyfert galaxy that shows evidence for 
reddening of its nonstellar continuum and emission lines, including those from 
the NLR. Given the morphological and kinematic evidence that the extended 
narrow-line region (ENLR) in a Seyfert galaxy is ionized gas in the galactic 
disk (Unger et al. 1987), we suggest that the dusty lukewarm absorber is a 
highly ionized component of the ENLR seen in absorption. 

The other type of absorber
(i.e., intrinsic to the nucleus) can be identified through 
absorption-line variability, high outflow velocities, and/or covering factors 
that are less than one. Thus, it is possible in many cases to use this 
information to separate the two types of absorbers, permitting more focused 
studies of either phenomenon. We note that it may not always be 
possible to identify the origin of a particular absorber; for example, a 
saturated absorber with low {\it radial} velocity, a covering factor of one, 
and no evidence for variability does not automatically fall into the dusty 
lukewarm absorber class (i.e., in the host galaxy). However, we suggest that 
evidence for reddening by a column sufficient to produce the absorption lines, 
plus evidence for covering of the NLR, does place it in this class.

One interesting question is: why don't all Seyfert 1 galaxies show UV 
absorption? Since nearly all Seyfert galaxies are spirals, we might expect a 
significant column of gas for even face-on objects.  Xilouris et al. 
(1999) find that a line-of-sight to the nucleus of a normal
spiral galaxy will show a column of a N$_H$ $\approx$ 5 x 10$^{20}$ 
cm$^{-2}$ at zero inclination, which should normally produce observable UV 
absorption lines (\S 3).
However, $\sim$40\% of Seyfert 1 galaxies show no UV absorption lines at all, 
including Ly$\alpha$ (Crenshaw et al. 1999). The answer may be that gas 
originally associated with the host galaxy is evacuated in the vicinity of the 
nucleus, (i.e., distances $\ltsim$ 100 pc) as suggested by de Zotti \& Gaskell 
(1985). This evacuation would occur at least within some polar angle relative 
to the axis of the accretion disk and/or putative torus (and within the 
ionization bicone through which we presumably view Seyfert 1 galaxies). 
Support for partial evacuation of the bicone comes through kinematic models of 
the outflowing gas in the NLR (Crenshaw et al. 2000). As we mentioned 
previously, another possibility for the lack of UV absorption is that the gas 
is highly ionized (U $>$ 3) close to the nucleus.

In many, if not all, Seyfert galaxies, there is clearly another source  of 
extinction at distances less than 100 pc from the nucleus (e.g., a dusty torus 
or dusty wind). Only a small fraction of Seyfert 2 galaxies can be 
attributed to viewing Seyfert 1 host galaxies at large inclination angles, due 
to the small range of inclination angles that would completely extinguish the 
nucleus. Furthermore, the source of extinction is not, in general, coplaner 
with 
the host galaxy, since Seyfert 2 galaxies are not seen 
preferentially edge-on (Kirhakos \& Steiner 1990). An important pursuit would 
be to 
determine the connection, if any, between reddening and UV absorption in the 
inner regions, since this would undoubtedly lead to a better understanding of 
the origin of the absorbers.  An interesting example is the
high-column (N$_H$ $=$ 2.7 x 10$^{21}$ cm$^{-2}$) absorber in NGC~4151, which 
is 
dust-free because it is located at a distance of only $\sim$0.03 pc (Kraemer 
et 
al. 2000), which is well within the dust sublimation radius for this Seyfert.
Studies of the reddening properties of other high-column UV absorbers would be 
helpful in determining if this is a common occurrence. Knowledge of the 
inclination angle of the accretion disk and/or torus with respect to our line 
of sight is also very important, and could be obtained, for example, from 
kinematic models of the NLR outflow (Crenshaw et al. 2000).

\acknowledgments

We acknowledge support from NASA grant NAG5-4103. We thank
Michael French for assistance with the {\it IUE} data. This research has made 
use of the NASA/IPAC Extragalactic Database (NED) which 
is operated by the Jet Propulsion Laboratory, California Institute of 
Technology, under contract with the National Aeronautics and Space 
Administration. This research has also made use of NASA's Astrophysics Data 
System Abstract Service.

\clearpage
\figcaption[fig1.ps]{UV color vs. inclination
angle of the host galaxy for the Seyfert 1s classified by their
continuum properties as Type A (unreddened) or B (reddened) by Ward et al.
(1987). Clearly, the Type B galaxies have redder UV colors and tend to be
more inclined.}
\label{fig1} 

\figcaption[fig2.ps]{UV color vs. inclination angle for the Seyfert 1s in our 
sample (see Table 1). Seyfert 1s with known or suspected dusty 
lukewarm absorbers are labeled.}
\label{fig2} 

\clearpage
\begin{deluxetable}{lccccccc}
\tablecolumns{8}
\footnotesize
\tablecaption{Inclination, Reddening, and Absorption Data for a 
Sample of Seyfert 1s
\label{tbl-3}}
\tablewidth{0pt}
\tablehead{
\colhead{Name} & \colhead{Type$^{a}$} &
\colhead{b/a$^{b}$} &
\colhead{UV} &
\colhead{Gal.} &
\colhead{Flux} &
\colhead{Flux} &
\colhead{Source}\\
\colhead{} & \colhead{} &
\colhead{} &
\colhead{abs?} &
\colhead{E(B-V)} &
\colhead{(FUV)$^{c}$} &
\colhead{(NUV)$^{d}$} &
\colhead{}
}
\startdata
Mrk 335      &      & 0.80  & No & 0.035 & 6.60   & 3.10 & {\it HST} \\
III Zw 2     & A      & 0.66 & Yes & 0.098 & 1.93   & 0.72 & {\it IUE} \\
WPVS 007     &      & 0.70$^{g}$ & Yes & 0.012 & 0.25   & 0.26 & {\it HST} \\
I Zw 1       &      & 0.88  & Yes & 0.065 & 1.19   & 1.13 & {\it HST}  \\
Ton S180     &      & 1.00$^{e}$ & No & 0.014 & 3.53   & 1.85 & {\it HST} \\
Mrk 1152     &       & 0.42$^{e}$  & & 0.021 & 0.33   & 0.24 & {\it IUE} \\
Fairall 9    & A      & 0.80  & No & 0.027 & 2.69   & 1.28 & {\it HST} \\
NGC 526a     & B      & 0.67  & & 0.028 &    & 0.18 & {\it IUE} \\
Mrk 590      &       & 0.97  & & 0.037 & 4.00   & 2.10 & {\it IUE} \\
NGC 931      & B      & 0.21 & & 0.096 & 0.16   & 0.77 & {\it IUE} \\
ESO 198-G24  &       & 0.75$^{f}$  & & 0.035  & 1.26   & 1.10 & {\it IUE}\\
NGC 1566     &      & 0.79    & No & 0.009 & 0.33   & 0.13 & {\it HST}\\
3C 120       & B      & 0.61$^{e}$ & & 0.297 & 0.70   & 0.60 & {\it IUE} \\
Akn 120      & A      & 0.67  & No & 0.128 & 3.41   & 2.09 & {\it HST}\\
MCG 8-11-11  & B      & 0.47  & Yes & 0.217   & 0.59  & 0.92 & {\it IUE}\\
H0557-385    & B      & 0.26$^{e}$  & & 0.043 &    & 0.15 & {\it IUE} \\
Mrk 6        &      & 0.43  &  & 0.136 & 0.10   & 0.10 & {\it IUE} \\
Mrk 79       & A      & 0.69 & & 0.071   & 2.06   & 0.64 & {\it IUE} \\
NGC 3227     & C      & 0.45 & Yes & 0.023  & 0.05   & 0.27 & {\it HST} \\
NGC 3516     &      & 0.75   & Yes & 0.042 & 2.82   & 1.91 & {\it HST} \\
NGC 3783     & A      & 0.80  & Yes  & 0.119  & 5.40   & 2.42 & {\it HST} \\
NGC 3786     &        & 0.50  & & 0.024 & 0.20   & 0.20 & {\it IUE} \\
NGC 4051     & C      & 0.68  & Yes & 0.013 & 1.96   & 1.21 & {\it IUE} \\
NGC 4151     & A      & 0.74  & Yes & 0.028 & 22.50  & 11.69 & {\it HST} \\
Mrk 766      &      & 0.85$^{e,h}$ & & 0.020 & 0.50   & 0.64 & {\it IUE} \\
Mrk 205      &      & 1.00$^{e}$  & No & 0.042 & 3.18   & 1.14 & {\it HST} \\
NGC 4593     & C      & 0.78  & & 0.025 & 1.92   & 1.07 & {\it IUE} \\
MGC-6-30-15  & B      & 0.58  & & 0.062 & 0.09   & 0.19 & {\it IUE} \\
IC 4329A     & B      & 0.16  & & 0.059 & 0.05   & 0.08 & {\it IUE} \\
Mrk 279      &      & 0.68  & Yes & 0.016 & 12.10   & 3.82 & {\it IUE} \\
NGC 5506     & B      & 0.21$^{e}$  & & 0.060  & 0.20   & 0.24 & {\it IUE} \\
NGC 5548     & A      & 0.83   & Yes & 0.020 & 3.41   & 1.90 & {\it HST}\\
 & & & & & & & \\
 & & & & & & & \\
 & & & & & & & \\
 & & & & & & & \\
 & & & & & & & \\
Mrk 1383     & A      & 0.83$^{e}$   & & 0.032 & 9.21   & 2.75 & {\it IUE}\\
Mrk 478      &      & 0.85   & No & 0.014 & 3.56   & 1.77 & {\it HST} \\
Mrk 841      & A      & 0.94$^{e}$  & & 0.030 & 4.50   & 1.46 & {\it IUE} \\
Mrk 290      &      & 0.86   & Yes & 0.015 & 4.83   & 1.22 & {\it IUE} \\
Mrk 493      &      & 0.75$^{e}$ & No & 0.025   & 0.66   & 0.37 & {\it HST} \\
ESO 103-G35  & B      & 0.20   & & 0.076 &    & 0.24 & {\it IUE}\\
ESO 141-G55  & A      & 0.65  & & 0.111 & 10.95   & 3.33 & {\it IUE}\\
NGC 6814     &      & 0.86$^{e}$  & & 0.183 & 0.69   &  & {\it IUE} \\
Mrk 509      & A      & 0.85 & Yes & 0.057 &  5.21   & 2.93 & {\it HST} \\
II Zw 136    &      & 0.41  & Yes & 0.044 & 3.03   & 1.28 & {\it IUE} \\
NGC 7213     & C      & 0.91$^{e}$  & & 0.015 & 3.83   & 1.55 & {\it IUE}\\
NGC 7314     & C      & 0.43$^{e}$  & & 0.021 &    & 0.50 & {\it IUE}\\
Akn 564      &      & 0.52  & Yes & 0.060 & 0.59   & 0.51 & {\it HST}  \\
MRC 2251-178 &      & 1.00$^{e}$  & & 0.039 & 3.89   & 1.03 & {\it IUE} \\
NGC 7469     & C      & 0.58 & Yes & 0.069 & 3.47   & 1.95 & {\it HST} \\
MCG-2-58-22  &       & 0.67  & & 0.042 & 16.06   & 4.25 & {\it IUE} \\
\tablenotetext{a}{A $=$ ``bare'' nuclear source, B $=$ reddened AGN,
and C $=$ galaxy-dominated (from Ward et al. [1987]).}
\tablenotetext{b}{minor/major axes ratios from de Zotti
\& Gaskell (1985), unless otherwise noted.}
\tablenotetext{c}{Flux at 1350 \AA, in units of 10$^{-14}$ ergs s$^{-1}$ 
cm$^{-2}$ 
\AA$^{-1}$.}
\tablenotetext{d}{Flux at 2200 \AA~ (for {\it HST} spectra), or 2400 \AA~ (for 
{\it IUE} spectra) in units of 10$^{-14}$ ergs s$^{-1}$ cm$^{-2}$ 
\AA$^{-1}$.}
\tablenotetext{e}{from Kirhakos \& Steiner (1990).}
\tablenotetext{f}{from NED.}
\tablenotetext{g}{Measured from Palomar Sky Survey image (this paper).}
\tablenotetext{h}{This galaxy appears to be much more inclined in the 
{\it HST}/WFPC2 continuum image (see Malkan, Gorjian, \& Tam 1998).}
\enddata
\end{deluxetable}


\clearpage
\vskip3.0in
\begin{figure}
\plotone{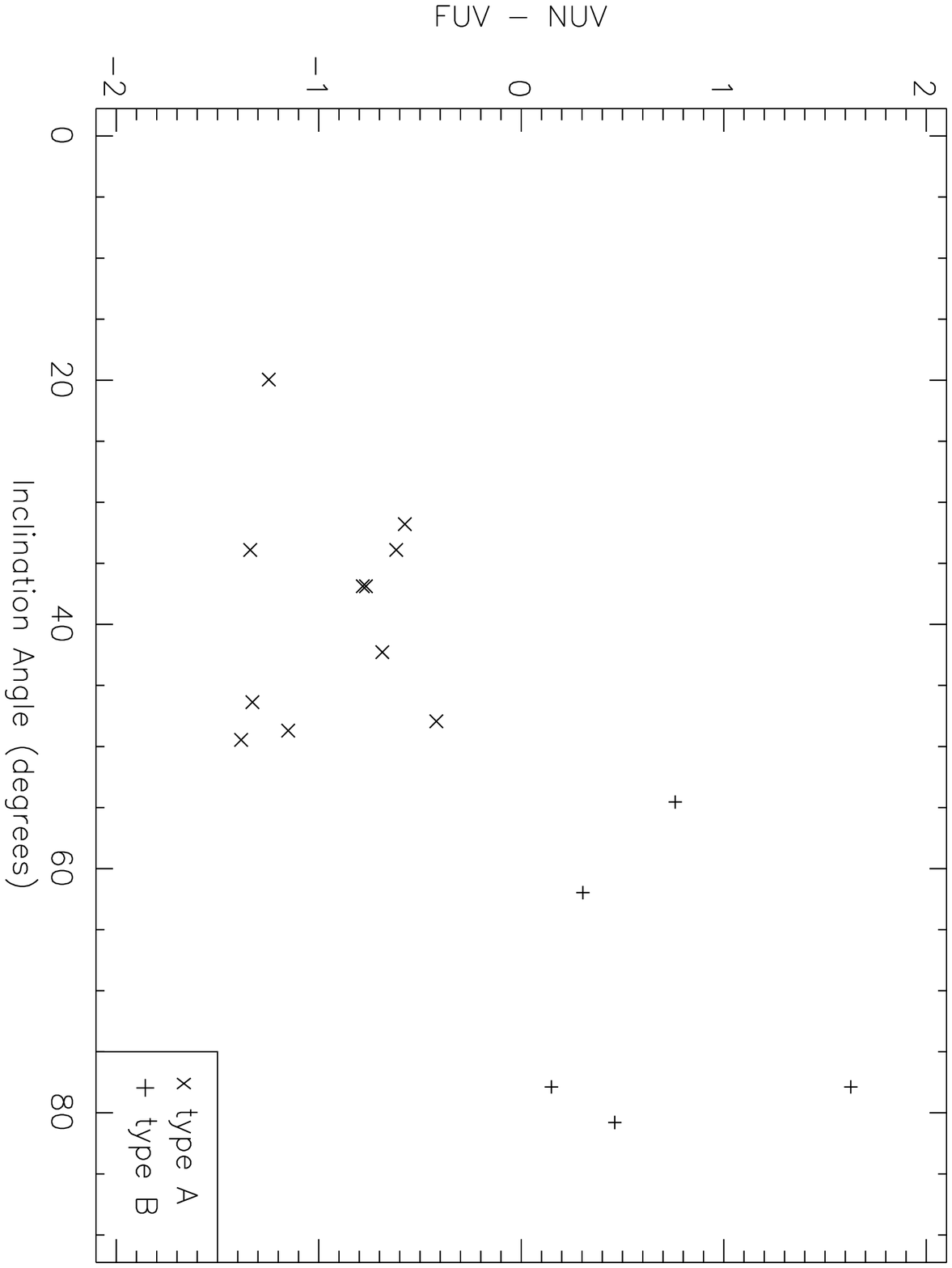}
\\Fig.~1.
\end{figure}

\clearpage
\vskip3.0in
\begin{figure}
\plotone{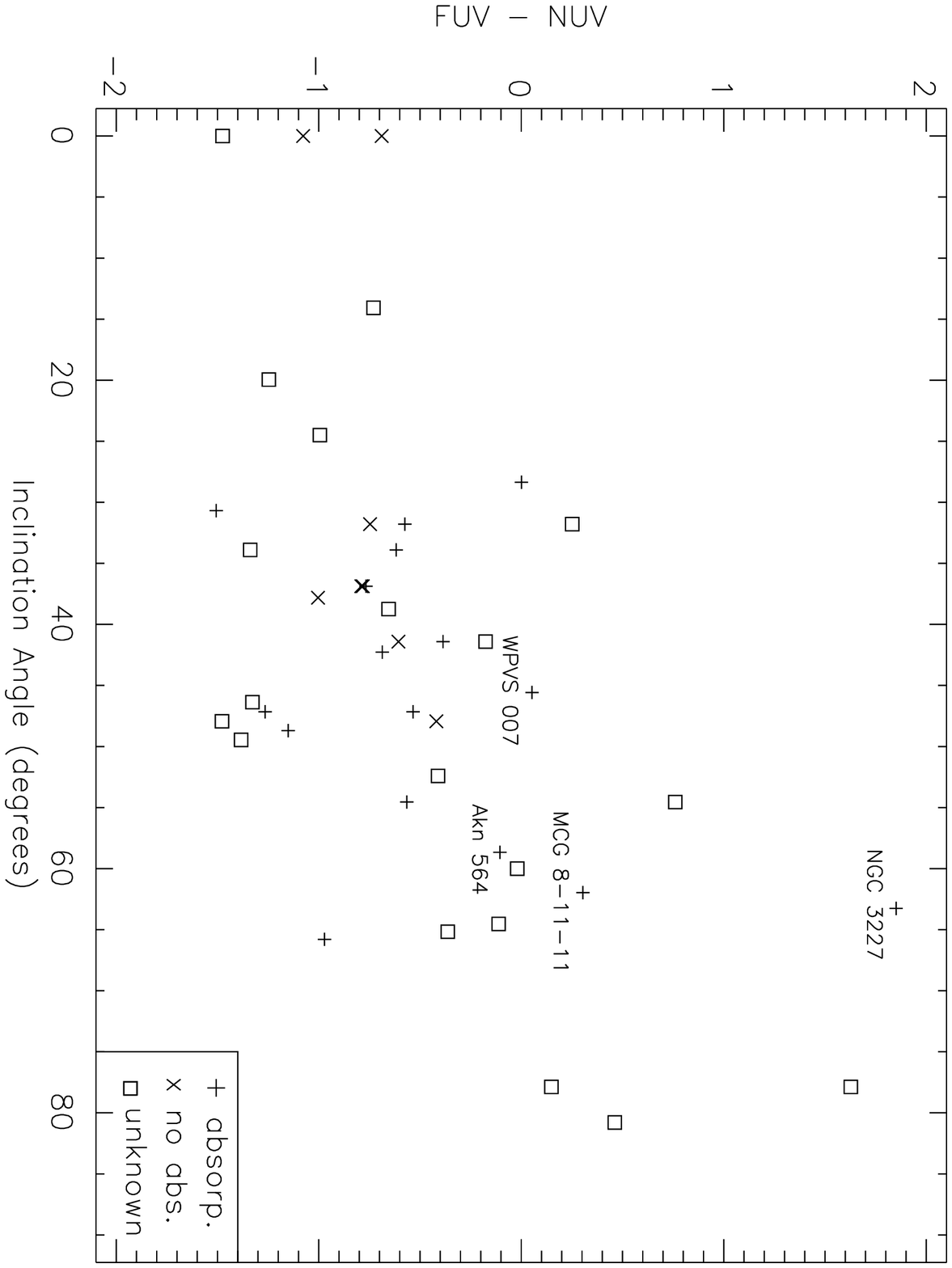}
\\Fig.~2.
\end{figure}

\end{document}